\documentclass{aa}

\usepackage[]{graphicx}
\usepackage[]{graphics}
\usepackage[]{psfig}
\usepackage[]{epsfig}
\usepackage[]{longtable}
\usepackage[]{times}
\include{epsf}

\begin{document}
\input{psfig.sty}


\title{The extended star formation history of $\omega$ Centauri\thanks{Based
on observations obtained at the European Southern Observatory, Chile (Observing
Programme 69.D--0172).}}

\author {Michael Hilker \inst{1} \and Andrea Kayser \inst{1} \and Tom Richtler
\inst{2} \and Philip Willemsen \inst{1}}

\offprints {M.~Hilker}
\mail{mhilker@astro.uni-bonn.de}

\institute{
Sternwarte der Universit\"at Bonn, Auf dem H\"ugel 71, 53121 Bonn, Germany
\and
Universidad de Concepci\'on, Departamento de F\'\i sica, Casilla 106-C,
Concepci\'on, Chile
}

\date{Received 29 April 2004/ Accepted 28 May 2004}

\titlerunning{The extended star formation history of $\omega$ Centauri}

\authorrunning{M.~Hilker et al.}

\abstract{For the first time, the abundances of a large sample of subgiant and
turn-off region stars in $\omega$ Centauri have been measured, the data base 
being medium resolution spectroscopy from FORS2 at the VLT.
Absolute iron abundances were derived for $\sim$400 member stars from newly
defined line indices with an accuracy of $\pm0.15$ dex. The abundances range
between $-2.2<$[Fe/H]$<-0.7$ dex, resembling the large metallicity spread
found for red giant branch stars. The combination of the spectroscopic results
with the location of the stars in the colour magnitude diagram has been used to
estimate ages for the individual stars. Whereas most of the metal-poor stars
are consistent with a single old stellar population, stars with abundances
higher than [Fe/H]$\simeq-1.3$ dex are younger. The total age spread in 
$\omega$ Cent is about 3 Gyr. The monotonically increasing age-metallicity 
relation seems to level off above [Fe/H]$\simeq-1.0$ dex. Whether the star 
formation in $\omega$ Cen occured continuously or rather episodically has to 
be shown by combining more accurate abundances with highest quality photometry.
}

\maketitle


\section{Introduction}
\label{intro}

$\omega$ Centauri is the most outstanding stellar cluster in our Milky Way
in many respects. It's the most massive and flattened cluster, and revolves
the Milky Way in a retrograde orbit, unlike most other Galactic
globular clusters. Many photometric and spectroscopic studies have
confirmed a wide spread in metallicity among its stars. This concerns all
observed elements (see the review by Smith \cite{smith04}).
It seems that the stars in $\omega$ Cen can be divided into three
main sub-populations:
1) a metal-poor population ($-2.0<$[Fe/H]$<-1.5$ dex), comprising
about 70\% of all stars; 2) an intermediate metallicity population
($-1.5<$[Fe/H]$<-0.9$ dex) with $\sim$25\% of
the stars; and 3) a distinct population ($\sim$5\%) of metal-rich stars
($-0.9<$[Fe/H]$<-0.6$ dex).
These sub-populations exhibit different behaviours in their spatial
distribution and kinematical properties. For references and a recent
summary on the
properties of $\omega$ Cen, see the contributions in the conference
proceedings by van~Leeuwen et al. (\cite{vanl02}).

Many ideas have been brought forward to explain $\omega$ Cen.
The most promising formation scenarios assume an extragalactic origin: 
it might be the nucleus of a
disrupted dwarf galaxy (first suggested by Zinnecker et al. \cite{zinne88}),
or the merger product of a super-cluster conglomerate that was created in an
interaction event of our galaxy with another star-forming galaxy (Fellhauer \&
Kroupa \cite{fell03a}).

An important parameter that can help to uncover the formation history of
$\omega$ Cen is the relative age of the different stellar populations.
Until now, only rough estimates of their ages
have been made, using broad and narrow band photometry
(Hilker \& Richtler \cite{hilk00b}, Rey
et al. \cite{rey04}, Hughes et al. \cite{hugh04}).
These studies suggest a time scale of chemical enrichment of up to 6 Gyr.
Hilker \& Richtler (\cite{hilk00b}, \cite{hilk02a}) argued for an extended
formation period of $\omega$ Cen in order to explain the tight correlation
between CN-band strengths and iron abundances.
Since nitrogen is predominantly provided by the debris of AGB stars and iron
by SNe II with greatly different evolutionary timescales, a natural
explanation would be multiple star formation events, triggered by gas
accretion from outside $\omega$ Cen and interrupted by long periods of
quiescence. Also the abundance pattern of s- and r-process elements demand
the contribution of low-mass AGB stars (Smith et al. \cite{smith00}).
Only the most metal-rich stars seem to be enriched by SNe Ia (Pancino et al.
\cite{panc02}).

In this paper, we present first results of a large spectroscopic survey,
dedicated to abundance measurements of subgiant branch (SGB) and main sequence
turn-off (MSTO) region stars. This is the most age-sensitive region
in the CMD. With the metallicity of a star in hand, its age can be estimated
by comparing its position in the CMD with appropriate isochrones.
We demonstrate that the suspected age spread among the stellar sub-population
in $\omega$ Cen definitely exists.

\section{Observations and data analysis}

The observations were performed in May 2002 with the
VLT/UT4 at Paranal (ESO), Chile. The instrument in use was the FORS2 camera
with the mask exchange unit MXU, and a 4$\times$4\,k MIT CCD attached. 

About 620 stars, selected from Str\"omgren photometry by Hilker \& Richtler
(\cite{hilk00b}), were observed through 11 slit masks in 5 fields around
$\omega$ Cen.
Additionally, spectra of 17 standard stars (Cayrel de Strobel et al. 
\cite{cayr01}) and MSTO and SGB stars in M55 have been
taken. Two grisms per mask were used: 1400V+18 and 600I+25 in second
order. The first grism has a dispersion of 0.62 \AA\,pix$^{-1}$ and
covers a wavelength range of 4560-5860 \AA, the specifications of the second
one are 0.58 \AA\,pix$^{-1}$ and 3690-4880 \AA. Together with the seeing and
a slit width of 1\arcsec, the resulting resolution is $\sim$2-2.5 \AA.

The CCD frames were processed with standard {\sc IRAF} routines. The 
signal-to-noise of the wavelength calibrated, rebinned (1\AA/pixel) spectra 
varied between 30 and 100 per pixel depending on the considered wavelength
range and the luminosity of the star.

Out of the 620 observed stars, 447 are, according to their radial velocities
and position in the CMD, SGB and MSTO region stars of $\omega$ Cen.
In M55, 38 member stars were observed in the MSTO/SGB region.

The determination of abundances was performed by measuring the 
pseudo-equivalent widths of several absorption lines.
For that, line indices have been defined analogous to the definition of Lick
indices, but with much smaller bandwidths. The uncertainties of the indices 
have been estimated from the noise in the continuum and sky spectra.
For the further analysis, the spectra have been classified by their quality
(good/bad signal-to-noise, right/wrong tracing, good/bad sky subtraction, 
etc.). About 430 spectra passed the quality check as good and best. 

The equivalent widths of the Balmer indices H$_\delta$ and H$_\gamma$ have been
combined to define an average Balmer index $\langle$H$\rangle$.
The iron abundances of the stars have been deduced from the average of
6 strong iron absorption lines between 4000 and 5300 \AA\,and a magnesium line 
at 5138 \AA. Prior to combination all indices have been scaled in such a way 
that the average line strength takes the value 1 at $\langle$H$\rangle= 3$.
This average index $\langle$Fe$\rangle$ has been used in the further analysis.
The effect of temperature and gravity $\log g$ on $\langle$Fe$\rangle$ has
been tested by simulating the measurements of $\langle$Fe$\rangle$ and
$\langle$H$\rangle$ on synthetic spectra with known parameters (taken from 
Bailer-Jones \cite{bail00}). Whereas the dependence of $\langle$Fe$\rangle$ 
on $\log g$ is small, $\langle$Fe$\rangle$ increases notably with decreasing 
$T_{\rm eff}$ within the temperature range of SGB stars.

However, the temperature effect alone cannot explain the large scatter in
$\langle$Fe$\rangle$ at all $\langle$H$\rangle$ values/colours (see Fig.~1).
This scatter is mostly due to the intrinsic iron abundance spread of the stars
in $\omega$ Cen.

\begin{figure}
\psfig{figure=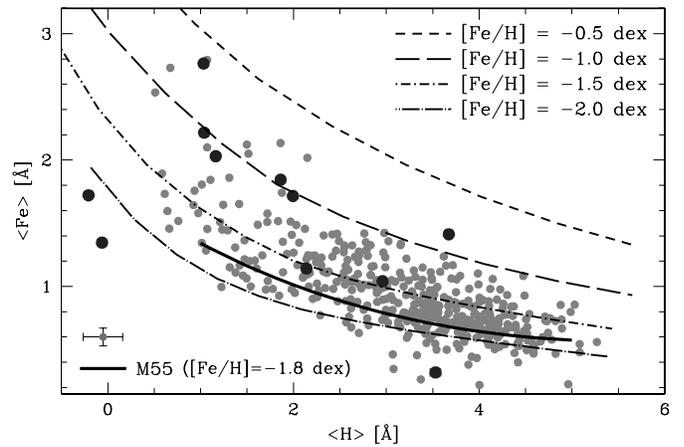,height=5.6cm,width=8.6cm
,bbllx=13mm,bblly=67mm,bburx=195mm,bbury=185mm}
\vspace{0.4cm}
\caption{\label{fig2} The iron index $\langle$Fe$\rangle$ is plotted vs. the
Balmer index $\langle$H$\rangle$ for 430 MSTO/SGB stars (grey dots). A typical
error is indicated on the left. Also shown are 11 standard stars
(bold dots) with $-2.7<$[Fe/H]$<-0.7$ dex. The bold curve marks a
fit to the position of 32 MSTO/SGB stars of M55. The other curves are 
iso-metallicity lines as measured from synthetic spectra. 
Their slope describes the temperature effect on $\langle$Fe$\rangle$ (about
5000 to 7000 K from left to right).
}
\end{figure} 

In order to find the absolute [Fe/H] values of the MSTO and SGB stars, their
distribution in the $\langle$H$\rangle$-$\langle$Fe$\rangle$ diagram (Fig.~1)
has been compared with that of the standard stars, the M55 stars,
and the iso-metallicity curves of the synthetic spectra. All these calibrators
are consistent with each other. 
To each star a [Fe/H] value has been assigned by establishing an analytical
relation between $\langle$H$\rangle$ and $\langle$Fe$\rangle$, based on the
data points of the calibrators. A polynomial of 4th order in both coordinates
gives an accurate fit with an rms of 0.02. The error
in [Fe/H] was propagated from the measurement errors of the line indices.
Typical errors are in the range $\pm$0.1-0.2 dex.
This is the so far highest accuracy in iron abundance determinations for 
MSTO/SGB stars in $\omega$ Cen. A direct comparison of the spectroscopic 
abundances of 165 MSTO stars with their determined Str\"omgren metallicities
(Hilker \& Richtler \cite{hilk00b}) reveals a scatter of about $\pm$0.5 dex in 
the photometrically deduced abundances at any [Fe/H] value. Moreover, the 
Str\"omgren colours have considerable photometric errors in the turn-off
domain and in addition are not only sensitive to iron but also to nitrogen.

Further details of the observations and data analysis will be given in Kayser
et al. (2004, in prep.).

\begin{figure*}
\psfig{figure=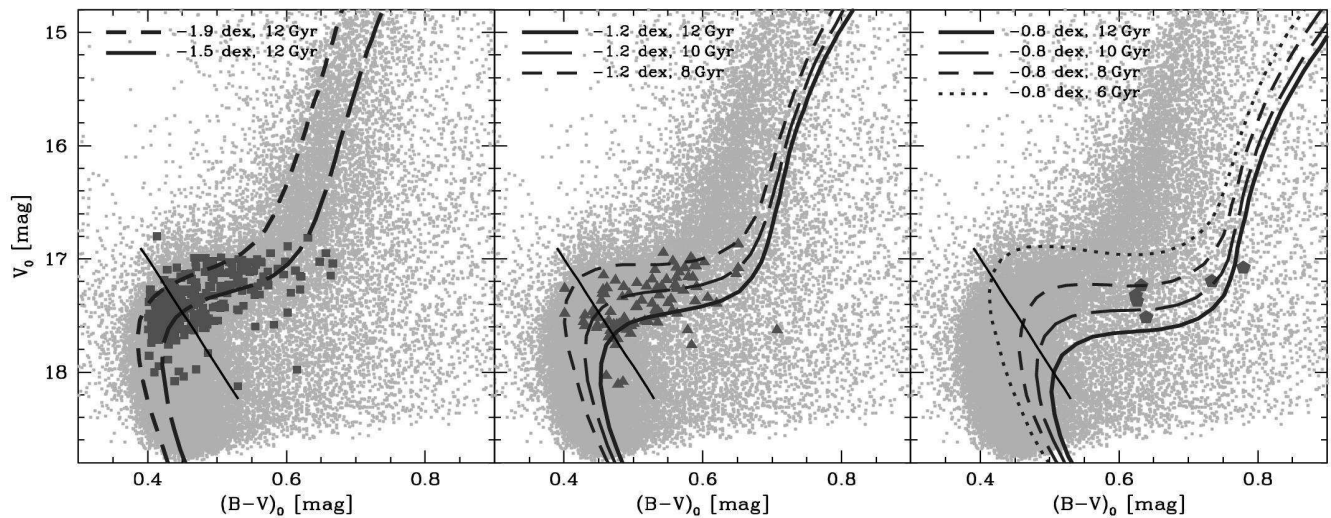,height=6.8cm,width=17.6cm}
\vspace{0.4cm}
\caption{\label{fig3} Comparison of the position of stars
of different abundances with Yonsei-Yale isochrones of different ages and
metallicities (as indicated in the upper left corners). The abundance ranges
for the highlighted stars are: left: $-2.0<$[Fe/H]$<-1.4$, middle: 
$-1.4<$[Fe/H]$<-1.0$, right: [Fe/H]$>-1.0$ dex. The diagonal
line indicates the selection criterium for age estimates (Fig.~3).
}
\end{figure*}

\section{Relating metallicities with ages}
\label{metal}

Having the [Fe/H] values for our sample stars, one can identify their
position in a CMD and fit them by appropriate isochrones.
For our purposes we used the $BV$ wide-field data by Rey et al. (\cite{rey04}),
corrected for reddening with $E_{B-V} = 0.11$ mag.
We used the Yonsei-Yale isochrones (Kim et al. \cite{kim02}) for the age
determination. The distance modulus to $\omega$ Cen was assumed to be $(m-M)_V
= 13.85$ mag (van~Leeuwen et al. \cite{vanl02}).

In Fig.~2, the distribution of stars for three metallicity ranges is shown.
The metal-poor population ($-2.0<$[Fe/H]$<-1.4$ dex) is mainly
located between the
isochrones of $-1.9$ and $-1.5$ dex and 12 Gyr. The $\alpha$-abundance is
assumed to be [$\alpha$/Fe]$=0.3$ dex. More metal-rich stars
($-1.4<$[Fe/H]$<-1.0$ dex) are not consistent with an isochrone of
12 Gyr, but rather scatter around isochrones that are up to 2
Gyr younger. Finally, the most metal-rich stars ([Fe/H]$>-1.0$ dex)
are distributed between isochrones of 9 to 11 Gyr when assuming a metallicity
of $-0.8$ dex and [$\alpha$/Fe]$=0.1$ dex (e.g. Pancino et al. \cite{panc02}).
From this analysis alone it is clear that the different sub-populations in
$\omega$ Cen do not share the same age.

To establish an age-metallicity relation, the age of each star was
extrapolated from an isochrone grid (Kim et al. \cite{kim02}) taking its 
metallicity as determined
from Fig.~1. The grid steps were 0.5 Gyr in age and 0.1 dex in metallicity.
The errors in the age determination for each star have been estimated by
taking isochrones of its maximum and minimum metallicity as defined by the
[Fe/H] error.

In Fig.~3, the age-metallicity relation for about 250 SGB stars 
(located redwards the diagonal line in Fig.~2) are shown. As a comparison
the histograms of 14 SGB stars of M55 are also shown. These stars are supposed 
to have a
single age ($\sim12.6\pm0.6$ Gyr) and metallicity ($-1.73\pm0.08$ dex), and
thus show the accuracy of the age and metallicity determination. 
Whereas the stars of $\omega$ Cen in the range $-2.1<$[Fe/H]$<-1.3$ dex
($\overline{\rm [Fe/H]}_{\rm mp} = -1.62\pm0.18$) might be consistent with a
single old age of $\sim12.1\pm1.4$ Gyr, the more metal-rich stars are 
significantly younger. The mean age of stars with $-1.3<$[Fe/H]$<-1.0$
($\overline{\rm [Fe/H]}_{\rm mr} = -1.19\pm0.07$) is $\sim10.5\pm1.7$ Gyr, and 
for the most metal-rich stars ([Fe/H]$>-1.0$, $\overline{\rm [Fe/H]}_{\rm mmr}
= -0.79\pm0.12$) $\sim9.4\pm0.9$ Gyr. In Table~1 the average ages of the 
different sub-populations in $\omega$ Cen are summarized in dependence of the
adopted distance modulus and reddening value.

\section{Discussion and Conclusions}

The abundance measurements of MSTO/SGB stars in $\omega$ Cen have shown that
the large metallicity spread seen for the RGB stars also exists for
MSTO stars, as expected. Although our spectroscopic sample may not be 
statistically complete due to selection effects, we see that the stellar 
population in the MSTO/SGB region is dominated by metal-poor stars (also 
consistent with the RGB results). There might exist a small
number of very metal-poor stars ([Fe/H]$<-1.9$). Still it has
to be shown whether these stars form a genuine cluster population or, for
example, have been accidentally caught by $\omega$ Cen.

The more metal-rich stars are definitely younger than the
dominant metal-poor population. This is the first time that the suspected age
spread has been confirmed by direct accurate abundance measurements in the
age-sensitive MSTO/SGB region. Although the absolute ages determined with
the Yonsei-Yale isochrones changes when using 
different reddening and/or distance modulus values, the overall trend of
metal-rich stars being younger than metal-poor ones does not change (see 
average ages of sub-populations in Table~1).

An interesting feature appears at the metallicity of 
[Fe/H]$\simeq-1.0$ dex. Whereas stars more metal-poor than $-1.0$ dex follow
a linearly increasing age-metallicity relation, stars above this metallicity
are not getting younger any more, but rather show the same age as stars 
around $-1.3$ dex. Due to the low number statistics it is not clear 
whether these stars are compatible with a single age or whether they have a 
significant age spread.

Whereas the existence of an age-metallicity relation is established, its
meaning and detailed properties still remain obscure. The width, for example, 
needs confirmation. At a metallicity of $-$1.5 dex, we find stars covering a 
large age interval. If this is true, self-enrichment of $\omega$ Cen must have 
had an extremely local character, which contradicts cluster-wide correlations 
between different elements. This adds to the difficulty of how a considerable 
amount of gas could have been retained for some Gyrs in spite of the enormous 
star formation activity. A more plausible scenario would be the accretion of 
gas from outside $\omega$ Cen, resulting in many star formation periods as 
sketched by Hilker \& Richtler (\cite{hilk00b}). 
The next step should be the 
investigation of abundance correlations among turn-off and main-sequence stars,
particularly the correlation between [Fe/H] and s-process elements (Smith 
\cite{smith04}). 
The Str\"omgren photometry indicates that already among the 
metal-poor population, the nitrogen enrichment was much faster than the 
enrichment of Ca, which is hard to explain with pure self-enrichment of 
$\omega$ Cen.

\begin{figure}
\psfig{figure=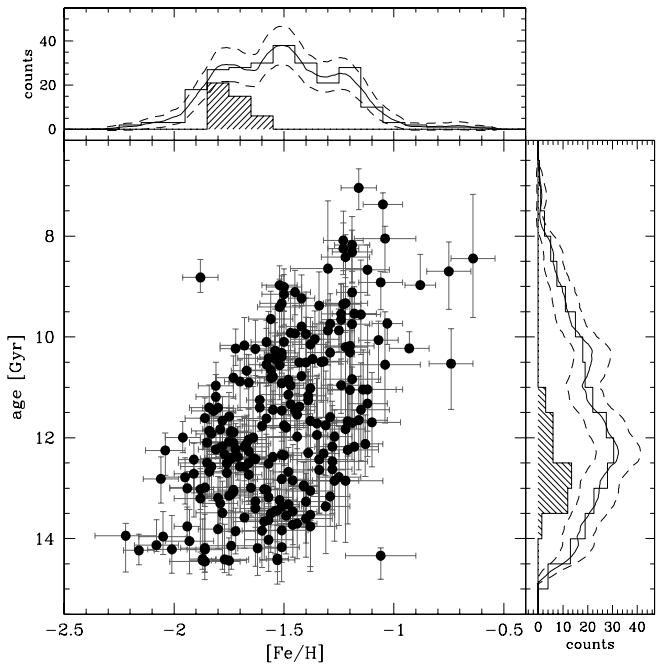,height=8.6cm,width=8.6cm
,bbllx=9mm,bblly=65mm,bburx=195mm,bbury=246mm}
\vspace{0.4cm}
\caption{\label{fig4} The age-metallicity relation for $\sim$250 SGB stars
(redwards the diagonal line in Fig.~2) with [Fe/H] errors less than 0.2 dex
and age errors less than 2 Gyr is shown.
The metallicity and age distributions are also plotted (as histograms and
binning independent density distributions). The hashed histograms represent
M55 stars (up-scaled by a factor of 3) which are supposed to have a single age
and metallicity ($-1.8$ dex).
}
\end{figure}

Recent deep photometric investigations of $\omega$ Cen with the HST revealed
the existence of multiple subgiant branches, turn-offs and a bifurcated main
sequence (Ferraro et al. \cite{ferro04}; Bedin et al. \cite{bedi04}).
Our results are in some contradiction to one of the suggestions by Ferraro et 
al. who propose that the SGB that seems to belong to the most metal-rich
population of the RGB ([Fe/H]$\simeq-0.6$) can be best fitted by an isochrone 
as old as the one for the metal-poor population. However, the exact shape of 
this SGB cannot be reproduced by their set of isochrones, perhaps indicating 
that some kind of abundance anomaly might be present.
Indeed, the reddest most metal-rich star (see Fig.~2) does not seem to be
compatible with any of the isochrones, and thus mimics an old age. 
We note that our sample of most metal-rich stars is very small (5 stars) and 
has an average metallicity of about $-0.8$ dex.
Unfortunately, our data have no overlap with the existing HST data.

Even more puzzling, the upper MS is bifurcated in such a way that about 25\%
of its stars lie bluewards to the bulk of MS stars (Bedin et al. 
\cite{bedi04}). This would normally be interpreted as a metal-poor stellar 
population. However, the ratio of metal-poor to metal-rich stars is just the 
opposite. If the very blue MS represents the more metal-rich population 
this would imply either an extraordinary high helium abundance of 
these stars or a distance gap between the populations. Bedin et al. suggest
that the intermediate metallicity population could be located about 1.6 kpc 
behind the metal-poor stars. This corresponds to 0.57 mag in distance modulus. 
Shifting the $-1.2$ dex isochrones (Fig.~2) by this amount to fainter 
magnitudes would decrease the ages of the intermediate metallicity stars by 
another 2-3 Gyr, thus even increasing the age spread within $\omega$ Cen. 
On the other hand, if the He content of these stars is very high,
this would shift its MS to bluer colours, and hardly would affect
the position of the SGB. Thus, the age estimates would not change.

\begin{table}[t!]
\caption{\label{tab1} Mean ages of different sub-populations in $\omega$ Cen. 
mp: metal-poor, mr: metal-rich, mmr: most metal-rich (see Sect.~3)}
\begin{tabular}[l]{ccccc}
\hline
 & & & & \\[-3mm]
$(m-M)_V$ & $E_{B-V}$ & $\overline{\rm age}_{\rm mp}$ & 
$\overline{\rm age}_{\rm mr}$ & $\overline{\rm age}_{\rm mmr}$ \\
$[$mag] & [mag] & [Gyr] & [Gyr] & [Gyr] \\
\hline
13.85 & 0.11 & $12.1\pm1.4$ & $10.5\pm1.7$ & $9.4\pm0.9$ \\
14.00 & 0.11 & $10.8\pm1.5$ & $9.0\pm1.5$ & $7.7\pm0.9$ \\
13.70 & 0.11 & $12.5\pm1.3$ & $11.1\pm1.6$ & $10.5\pm1.2$ \\
13.85 & 0.13 & $11.4\pm1.5$ & $9.8\pm1.6$ & $8.6\pm1.1$ \\
13.85 & 0.09 & $12.7\pm1.3$ & $11.3\pm1.6$ & $10.9\pm1.4$ \\
\hline
\end{tabular}
\end{table}

Although an age spread among the different populations in $\omega$ Cen has been
confirmed by our data and an age-metallicity relation has been established, the
population puzzle in $\omega$ Cen needs further observational and theoretical 
input.
With the new photometric data available (HST/VLT), we now have the possibility 
to accurately select SGB and upper MS stars from a certain sub-population for 
further spectroscopic analysis.

\acknowledgements
The authors are very grateful to S.-C. Rey for providing the $BV$ data.
T.R. acknowledges support by the FONDAP Center for Astrophysics, Conicyt 
15010003. We also thank C.A.L. Bailer-Jones for calculating synthetic spectra
at our request, and thanks to the anonymous refere for his useful comments.

\enddocument